\documentclass{iopjournal}
\usepackage{amsfonts}
\usepackage{xcolor}

\begin{document}

\articletype{Paper}

\title{Machine Learning for neutron source distributions}

\author{José Ignacio Robledo$^{1,2,*}$\orcid{0000-0000-0000-0000},  Norberto Schmidt$^{1}$\orcid{0000-0003-3600-1446}, Klaus Lieutenant$^{1}$\orcid{0000-0001-8278-9401}, Jingjing Li$^{1}$, Stefan Kesselheim$^{2}$\orcid{0000-0003-0940-5752}, and Paul Zakalek$^{1}$\orcid{0000-0002-6497-9604}}

\affil{$^{1}$ \quad Jülich Centre for Neutron Science (JCNS-2), Forschungszentrum Jülich, Jülich, Germany}

\affil{$^{2}$ \quad Jülich Supercomputing Centre (JSC), Forschungszentrum Jülich, Jülich, Germany}

\affil{$^*$Author to whom any correspondence should be addressed.}

\email{j.robledo@fz-juelich.de}


\begin{abstract}
In light of the recent advancements in machine learning, we propose a novel approach to neutron source distribution estimation through the utilisation of probabilistic generative models. The estimation is based on a Monte Carlo particle list, which is only required during the training stage of the machine learning model. Once the source distribution has been learned, the model is independent of the original particle list, allowing for further sampling in an efficient, rapid, and memory-costless manner. The performance of various generative models is evaluated, including a variational autoencoder, a normalizing flow, a generative adversarial network, and a denoising diffusion model. These approaches are then compared to existing source distribution estimations, and the advantages and disadvantages of each approach are discussed. The results demonstrate that source distributions can be modeled through the use of probabilistic generative models, which paves the way for further advancements in this field.
\end{abstract}

\section{Introduction}

In the field of neutronics, there is a clear necessity for an accurate charachterization of the distributions of neutron phase-space variables at various points within the neutron source. These distributions are intrinsically high-dimensional, strongly correlated, and shaped by complex nuclear interactions occurring inside the source. The simulation of neutron sources is frequently conducted using software such as MCNP \cite{forster2006mcnp}, PHITS \cite{niita2006phits}, OpenMC \cite{romano2015openmc}, amongst others. While highly accurate, these simulations are typically time-consuming, particularly when high statistics are required to observe rare events, as they allow for a variety of nuclear processes including the production of neutrons. A standard output of such simulations is a lists of particles that traverse a specified surface, which can be stored in standardised formats such as the Monte Carlo Particle List (MCPL) \cite{kittelmann2017monte} format. This MCPL file becomes a finite sample of the multivariate phase-space variable distribution at a specific surface and time of the simulation, and can be reused as a source description for downstream neutron ray-tracing simulations in instrument design packages such as Vitess \cite{mc-neutron1}, McStas \cite{mcstas}, McVine \cite{lin2016mcvine}, Prompt \cite{mc-neutron4}. However, the direct reuse of MCPL files has well-known limitations: undersampling, statistical noise, and biases present in the original simulation are propagated unchanged often leading to artificial structures (such as hot spots) and poor convergence in subsequent simulations. 
Another common approach to estimate the source distribution is to fit analytical or semi-analytical models to individual source variables in the MCPL file \cite{willendrup2020mcstas}. While computationally efficient, this approach typically assumes independence or weak coupling between variables and may lead to the loss of meaningful correlations between phase-space variables at the surface under consideration. Preserving these correlations is essential for realistic neutron transport, particularly for modern high-brilliance sources where phase-space structure directly impacts instrument performance.

Recent machine learning (ML) approaches have already shown success in overcoming these limitations in the field of neutronics, like KDSource \cite{KDSOURCE}, a python package for the estimation of source distributions by means of adaptive multivariate Kernel Density Estimation, or the use of neural networks to learn the scattering law $S(\alpha, \beta)$ conditioned to the neutron energy \cite{refId0, forget2022normalizing}
. Also, there have been attempts on using Generative Adversarial Networks in other domains, such as x-ray Monte Carlo simulation for medical applications \cite{Sarrut_2019}. However, their practical deployment as neutron source representations in Monte Carlo workflows, their comparative performance across ML model architectures, and their validation against experimental data remain largely unexplored.

In this work, we address this gap by presenting a systematic, application-driven study of modern generative models to model source distributions, with the explicit goal of replacing raw particle lists as source descriptions in neutron ray-tracing simulations. For this, we compare four distinct models: Variational Autoencoders (VAE), Normalizing Flows (NF), Diffusion models (DM) and Generative Adversarial Networks (GAN). The selection of these models reflects a deliberate comparison across fundamentally different generative paradigms, rather than an attempt to optimize a single architecture. We evaluate these approaches against two baselines: direct use of the MCPL file as input to tracing simulations and to KDSource as a black-box source distribution estimator. In this way, we demonstrate that mondern generative approaches can be successfully integrated into neutron Monte Carlo pipelines and yield physically meaningful results across different problems.

This work goes beyond the abstract ability of generative models to approximate complex phase-space distributions by focusing on their concrete realization, integration, and validation within neutron Monte Carlo simulation pipelines. Unlike analytical estimation, fitting procedures, or KDSource, the proposed generative models learn latent spaces that, when sampled, produce accurate representations of the original distribution. Once trained, these models operate independently of the original particle list, making them well-suited for integration into Monte Carlo software for source representations. To this end, we have developed a dedicated source module in the MC software Vitess \cite{mc-neutron1} called \texttt{source\_AI}, which enables the direct use of trained generative models as neutron sources, and use it to analyze two datasets: an MCPL file used in the Technical Design Report (TDR) of the High Brilliance Source (HBS) project  \cite{HBS2023} of the Jülich Centre for Neutron Science (JCNS) at Forschungszentrum Jülich (FZJ) and an MCPL file generated from a PHITS simulation that was benchmarked against the experimental measurements performed in the JULIC Neutron Platform during 2023 \cite{schmidt_2025}. To our knowledge, this constitutes the first comparative study in which modern generative ML models for neutron source modeling are integrated into MC software and validated against experimental data.

\section{Methods}\label{sec:methods}

We want to model the joint probability distribution $p(x,y,\theta,\varphi,p,t,E)$ of the phase-space variables of the neutrons arriving a surface where an MCPL file is recorded.
We propose a comparative analysis of four distinct probabilistic generative models (PGMs) for learning the multivariate distribution of phase-space variables of neutrons generated in a MC simulation of a source. 
The models that will be compared are described in section \ref{sec:models}. After training, they can be used to generate new neutrons that follow the original MCPL phase-space variables multivariate distribution, most importantly preserving correlations between variables. All models consist of a set of floating point parameters connected by weights in an architecture through simple operations. 

  \begin{figure}
      \centering
      \includegraphics[width=0.48\linewidth]{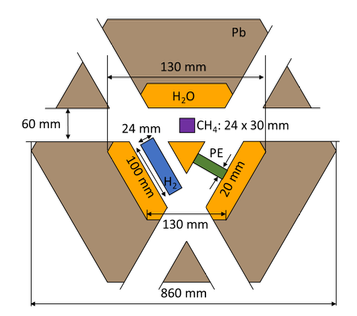}\hfill
      \includegraphics[width=0.48\linewidth]{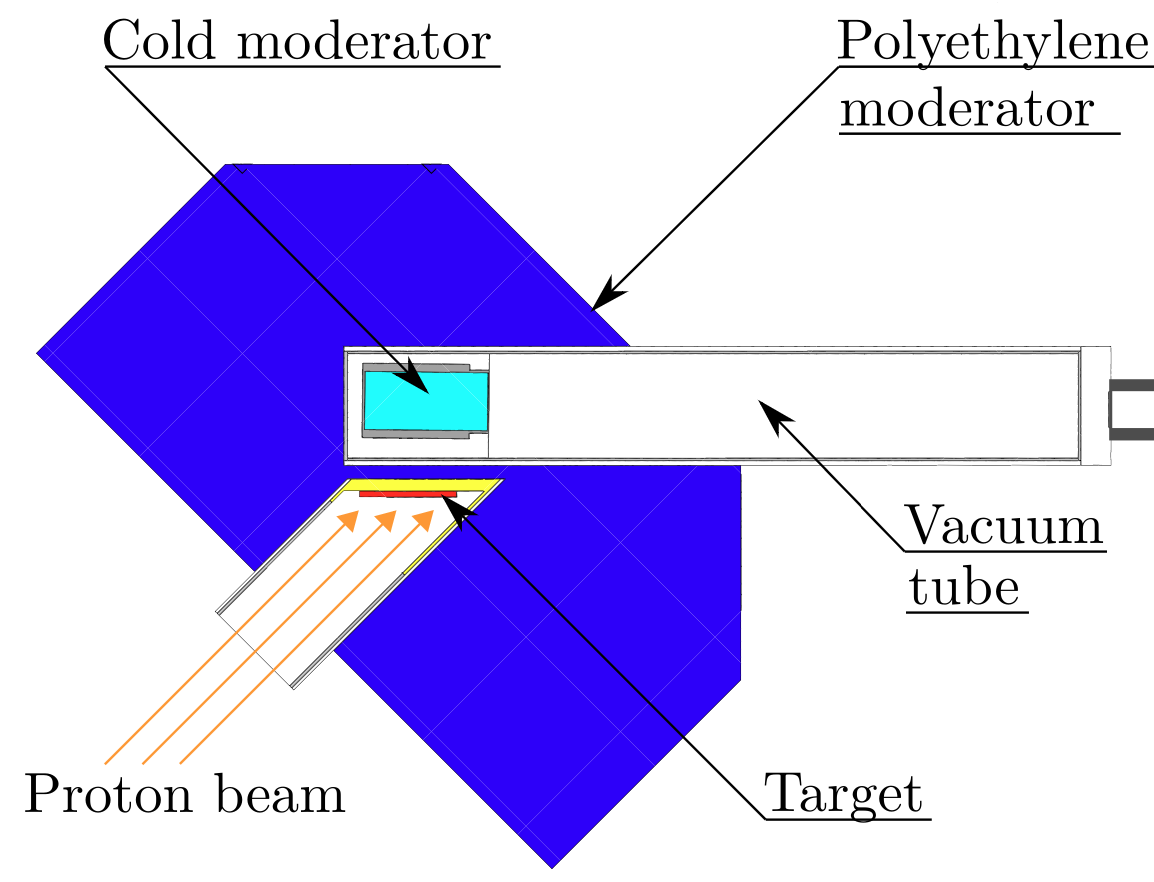}
      \caption{Left: Schematic of the Target-Moderator-Assembly used for the PHITS simulations. Copied from \cite{2023-Zakalek.} under the terms of the \href{https://creativecommons.org/licenses/by/4.0/}{Creative Commons Attribution License 4.0}. Right: Schematic of experimental setup adopted from \cite{schmidt_2025} with accordance to terms in \href{https://creativecommons.org/licenses/by/4.0/}{Creative Commons Attribution License 4.0}.}
      \label{fig:experiment}
  \end{figure}

To showcase the versatility of generative models in representing complex multivariate distributions, we train all models in two different datasets. The first dataset has been generated by means of MC simulations of the HBS para-hydrogen moderator as published in the TDR of the HBS project \cite{HBS2023}. This dataset was generated using PHITS, and is an MCPL file of $6\times10^{5}$ neutrons crossing a surface that is normal to one of the extraction channel, and was used afterwards as an input for Vitess simulations of a beamline inserted in the extraction channels. A schematic of the moderator is shown in Fig. \ref{fig:experiment}.  The second dataset consists on a benchmarked simulation of the target moderator station of the JULIC neutron facility mounted at Forschungszentrum J\"ulich during 2023, and already studied by means of KDSource for source distribution estimation by Schmidt et al. \cite{schmidt_2025}. This dataset consists on $1\times10^5$ neutrons, also crossing a surface normal to an extraction channel. The interest in this dataset lies in that we can compare the outputs of the MC tracing software Vitess using the artificial intelligence (AI) sources presented in this work with measured experimental data and with the KDSource approach. 

In what follows, we give a brief explanation of the structure of the two datasets used, and to the artificial intelligence models trained to learn the probability distributions of the data, as well as the training process and the tools created for sampling.

\subsection{Monte Carlo Simulation}

We focus on neutron phase-space distribution modeling, but Monte Carlo Particle List (MCPL) files can be used to train generative models in many different applications and of different type of particles. For detailed information on MCPL files see Kittelmann et al. \cite{kittelmann2017monte}. In a brief summary, they are lists of particles crossing specific surfaces and can be generated when running Monte Carlo simulations. The phase-space variables stored are the position ($x$, $y$, $z$ in $cm$), direction of movement (unit vector, $u_x$, $u_y$, $u_z$), time (ms), energy (MeV), weight (in case of variance reduction), type of particle (neutron, photon, etc. For codification refer to Ref. \cite{kittelmann2017monte}) and polarization (unit vector, $s_x$, $s_y$, and $s_z$). The focus of this work is on MCPL files generated in neutron transport codes \cite{romano2015openmc, forster2006mcnp, niita2006phits} that are to be used as source description in neutron tracing simulation software \cite{mc-neutron1, mcstas, lin2016mcvine, mc-neutron4}. Rather than importing the MCPL file (which can be in the order of several GB large, or even larger), we attempt to learn its distribution by training PGMs on it. This allows fast and simple generation of MCPL files that would have been produced in different iterations of the neutron transport simulation without the need of performing new time consuming simulations, and also straightforward sharing of source models that enable on-the-flight sampling. This is due to the fact that PGMs can be stored in a few kilobyte-sized file as a result of the simplicity of their architectures. 

\subsection{Datasets}\label{sec:datasets}

We used two MCPL files, which we will further on call TDR and Benchmark datasets. The TDR dataset is the result of a PHITS simulation of the moderator in one of the target stations of the High Brilliance Source (HBS) project, which was described and used in its Technical Design Report (TDR) \cite{Brckel:1016731}. Here, the 676.558 neutrons that cross a surface set normal to the $x$ direction at the entrance of an extraction channel are stored in an MCPL file. The relevant variables for this dataset are the positions $y$ and $z$ (measured in centimeters), the polar and azimuthal angles of the velocity direction $\theta$ and $\varphi$ (measured in degree), the particle weight $p$, the time of flight $t$ (in milliseconds), and the energy $E$ (in MeV). The angles $\theta$ and $\varphi$ are the polar coordinate angles of the velocity vector, which can be obtained from the MCPL direction variables $u_x$, $u_y$, and $u_z$ by $\theta=\arccos{u_z}$ and $\varphi=\arctan{\left(\frac{u_y}{u_x}\right)}$. The velocity magnitude is encoded in the energy.
Due to the large spread of thermal and fast neutrons, the decadic logarithm is applied to the variables $t$ and $E$, a common procedure in Machine Learning. For a better comprehension on the variance-covariance structure of this dataset, a pair-plot figure showing 1D and 2D histograms of all variables can be found in the Supplementary Material. 

The second dataset used to show PGMs ability to estimate complex multivariate phase-space variables distributions belongs to a simulation of the experiments performed in 2023 at the JULIC Neutron Platform of FZJ \cite{refId15}. The experimental setup and measurements are thoroughly explained in ref. \cite{schmidt_2025}. The resulting MCPL file, consisting of 118.219 neutrons, was used in a Vitess simulation to benchmark the experimental spectra measured at the end of a neutron guide, therefore we call it the Benchmark dataset. For this source simulation, another strategy was used in terms of variance reduction, therefore all Monte-Carlo weights of particles reaching the surface where the MCPL file was generated were equal to 1. This reduced the dimensionality of the problem to 6 dimensions, ($y$  [cm], $z$  [cm], $\theta$  [deg], $\varphi$  [deg], $\log_{10}(t)$  [ms], $\log_{10}(E)$  [MeV]). A figure showing the multivariate distribution structure can be found in the Supplementary Material.  

Before training on both datasets, all variables were normalized by means of $MinMax$ normalization,
\begin{equation}
    f_{MinMax}(x) = \frac{x - \min(x)}{\max(x) - \min(x)} = x_{norm}.
\end{equation}
This is a common procedure to facilitate the training process of machine learning (ML) models. Since the PGMs will be used to generate new neutron data, it is essential to store the minimum and maximum values of the training datasets together with the model. This ensures that the PGMs produce variables within the normalized interval [0,1], which can subsequently be accurately rescaled to the original phase-space domain by
\begin{equation}
    f^{-1}_{MinMax}(\hat{x}_{norm}) = \hat{x}_{norm} \left(\max(x) - \min(x)\right) + \min(x) = \hat{x},
\end{equation}
where $\hat{x}_{norm}$ is the vector generated by the PGM, and $\hat{x}$ the generated phase-space variables of a new neutron.

The two datasets considered exhibit distinct characteristics that pose different challenges for accurate modeling, providing means to evaluate the capability of the PGMs to capture such features across varying dimensionalities. The TDR dataset presents a more complex distribution, making it a particularly demanding modeling task. In contrast, the Benchmark dataset is less complex but offers the advantage of readily available experimental data for direct comparison.  

\subsection{Models}\label{sec:models}

This section provides a concise introduction to the different PGMs used in this work for readers not familiarized with generative models \cite{genaireview}. For those already acquainted, we suggest proceeding directly to section \ref{sec:implementation} where we explain the implementation of the models in our work. All AI models discussed attempt to learn the multivariate probability distribution $p(x)$ from samples of it, or directly learn to generate new samples that follow the original distribution. The set of available samples $\{x_i\}$  is referred to as the training dataset. We present four distinct models: NFs, VAEs, GANs, and DMs. The fundamental idea of each approach is explained in the following subsections along with a short summary of their advantages and disadvantages. All schematics in this section illustrate an input vector $x = (y, z, \varphi, \theta, p, t, E)$ and a corresponding output vector $\hat{x} = (\hat{y}, \hat{z}, \hat{\varphi}, \hat{p}, \hat{t}, \hat{E})$ both of dimension seven because they use the implementation on the TDR dataset described in section \ref{sec:datasets} as an example. For the benchmark dataset the architectures are analogous, except that the weight vector $p$ (and $\hat{p}$) is omitted from the input and output vectors.

\subsubsection{Normalizing Flow (NF)}\label{sec:NF}
\begin{figure}[h!]
    \centering
    \includegraphics[width=0.7\linewidth]{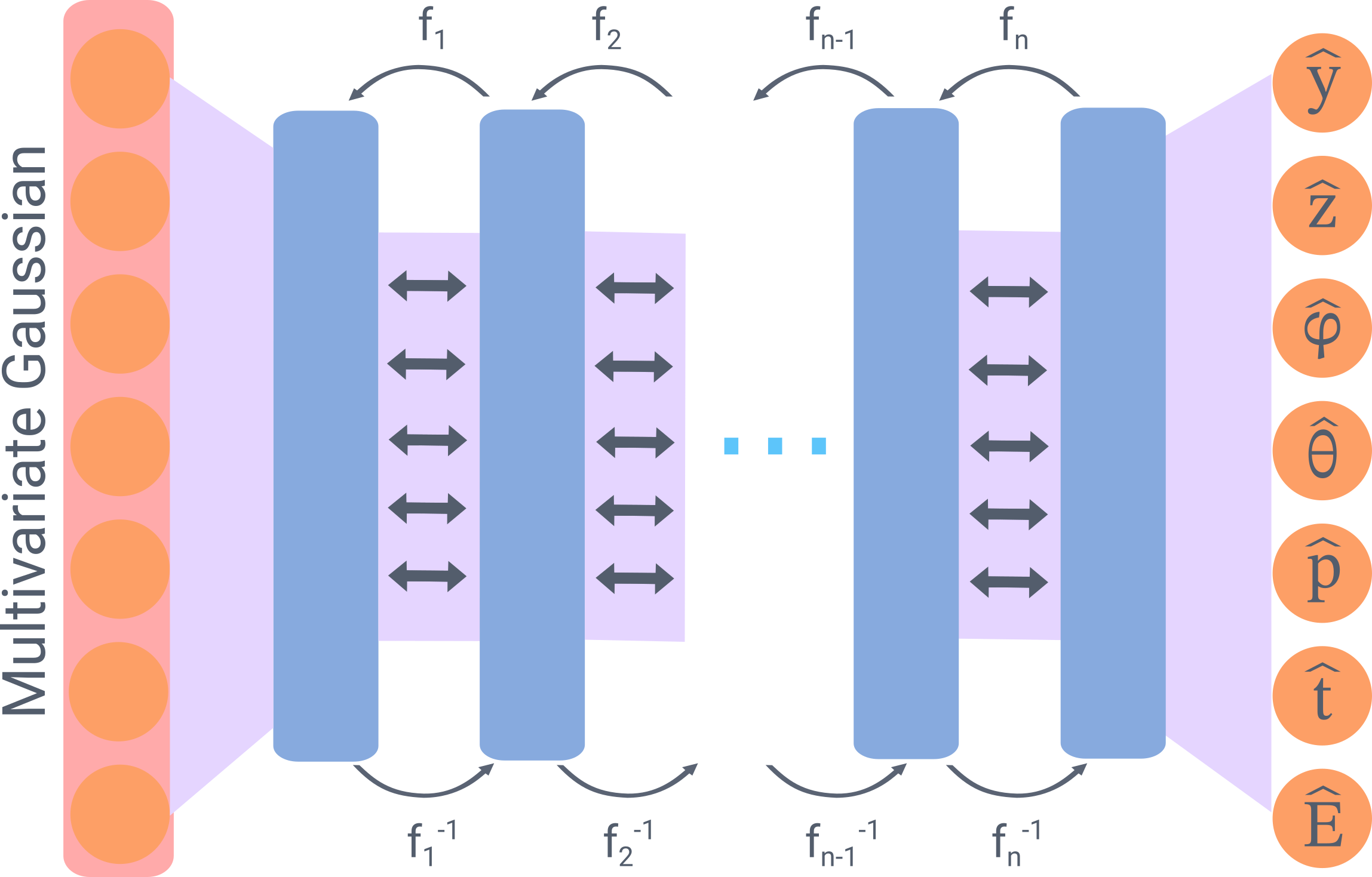}
    \caption{Conceptual schematic of a Normalizing Flow model. Blue blocks represent the invertible transformations of data $f_i(x)$, which gradually shape the phase-space volume to that of a multivariate gaussian of the same dimensions as the input dimension. Invertability allows the probability flow to go from the multivariate gaussian space, where it is easy to sample, to the phase-space.
    }
    \label{fig:schematic_nflow}
\end{figure}

These models are a class of generative models that learn a sequence of invertible transformations $f_i$ mapping between a complex data distribution $p(x)$ and a simple, well-known reference distribution $p(z)$. By successively applying these functions, samples from the data space $x$ can be transformed into the reference space $z$ and vice versa (see Fig.~\ref{fig:schematic_nflow}). The most common choice for the reference distribution is a multivariate Gaussian with zero mean and unit variance, with dimensionality chosen to match that of the data.  

The key property of NFs is invertibility: each transformation can be run forward (data $\to$ latent) or backward (latent $\to$ data). This makes it possible to both, evaluate likelihoods and generate new samples. Sampling proceeds by drawing from the simple distribution (e.g.\ Gaussian) and applying the learned transformations to obtain realistic data. The mapping functions are typically parameterized by invertible neural networks, where each layer $f_i$ is designed to be expressive yet computationally efficient to invert. The depth of the flow (i..e. the number of transformations) controls how well the model can approximate complex data distributions.  

A practical consideration is that flows rely on smooth invertible functions. When the target distribution has sharp features, NFs may approximate them poorly unless preprocessing or specialized architectures are applied. For instance, applying known invertible transformations that reduce sharpness before training can improve performance; alternatively, deeper or more expressive flows may be required. The trainable parameters of the flow layers are optimized by maximizing the likelihood of the observed data under the model, which corresponds to using the forward transformation into Gaussian space during training.

A variety of layer designs exist for NFs. Two popular families are Coupling Flows \cite{coupling2014} and Masked Autoregressive Flows (MAF) \cite{MAF2016}. Coupling flows are relatively simple, easy to invert, and efficient for sampling, while MAFs generally achieve higher expressivity at the cost of slower sampling due to their autoregressive structure. A more detailed comparison can be found in Ref.~\cite{sym16080942}.

In our study, we employ NFs with 10 flow layers. We experiment with two architectures: one based on coupling flows and one using only MAF layers, referred to as Coupling Flow and MAF respectively. For the TDR and Benchmark datasets, the reference distribution is chosen as a multivariate Gaussian of dimension seven and six, respectively. Since the MinMax-normalized values in the MCPL file lie between 0 and 1, we include a final Sigmoid layer ($\sigma(x) = (1 + e^{-x})^{-1}$), which is invertible and thus compatible with the NF framework, to force the output to be in the [0,1] interval (hard constraint).

\subsubsection{Variational Autoencoder (VAE)}\label{sec:VAE}

They are generative models that combine deep learning with variational Bayesian inference to learn latent representations of data. They have been widely applied in diverse domains~\cite{singh2021overview}. 

\begin{figure}[h!]
    \centering
    \includegraphics[width=0.7\linewidth]{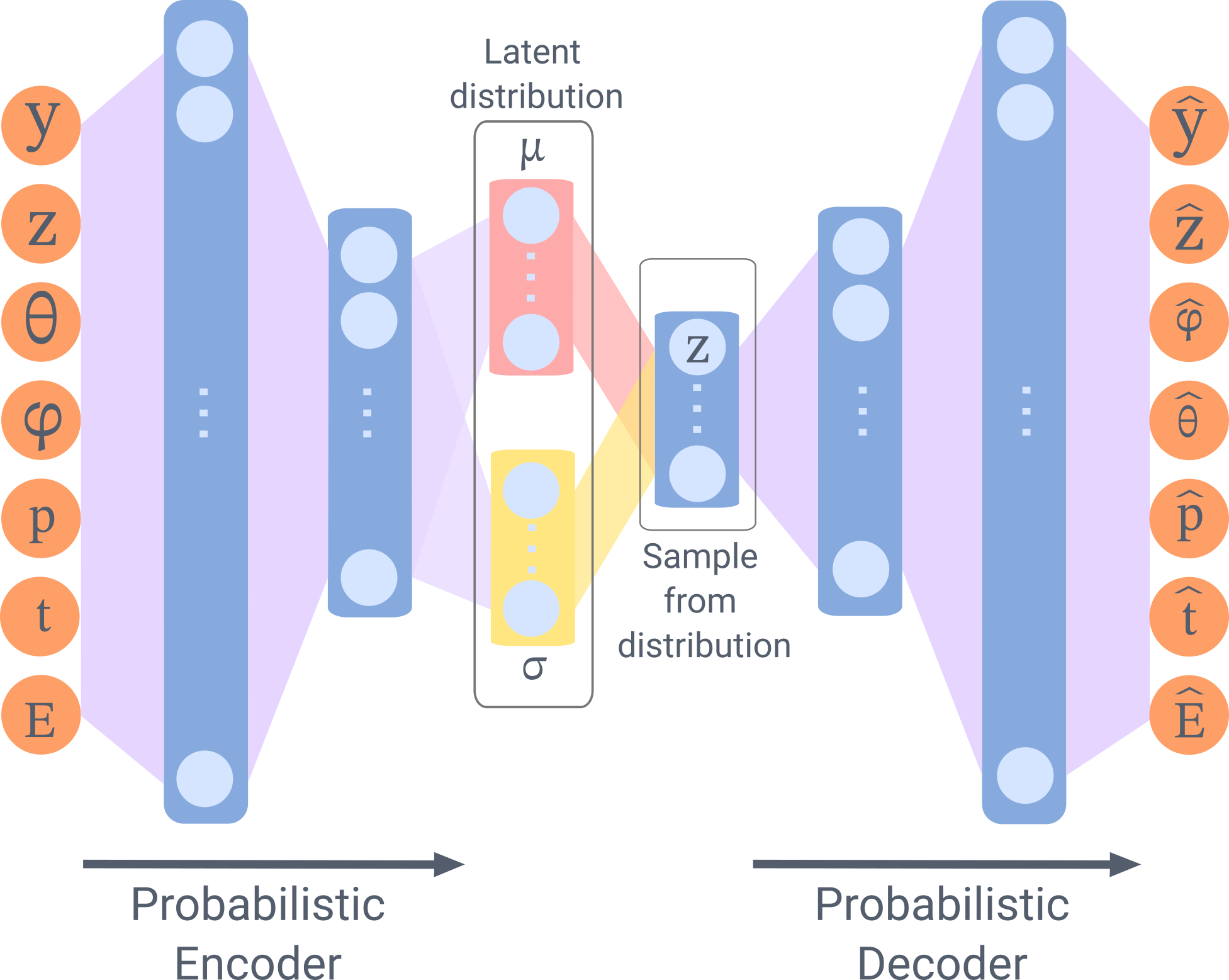}
    \caption{Conceptual schematic of a VAE model. Blue boxes represent layers that are fully connected by learnable parameters (purple). Latent space is generated by learning the mean (orange) and standard deviation (yellow) vectors of dimension $d_{latent}$. 
    }
    \label{fig:schematic_VAE}
\end{figure}

Conceptually, a VAE consists of two neural networks: an \emph{encoder} and a \emph{decoder}. The encoder maps input data $x$ to a distribution over latent variables $z$. In practice, the encoder outputs the parameters of this distribution, typically the mean $\mu$ and variance $\sigma^2$ of a multivariate Gaussian. A latent vector $z$ is then sampled from this distribution and passed through the decoder, which attempts to reconstruct the original input. A schematic overview is given in Fig.~\ref{fig:schematic_VAE}.  

The key idea of a VAE lies in its loss function, which balances two objectives. The first term encourages accurate reconstruction of the input (commonly via mean squared error between $x$ and its reconstruction $\hat{x}$). The second term is a Kullback--Leibler (KL) divergence regularizer that pushes the learned latent distribution towards a chosen prior, usually a standard multivariate Gaussian. This regularization ensures that the latent space is smooth and continuous, enabling the generation of new, realistic samples by randomly sampling $z$ from the prior distribution. For a broader review of autoencoders and their variants, see Ref.~\cite{berahmand2024autoencoders}.  

Training VAEs requires the \emph{reparameterization trick}~\cite{train_vae}, which makes sampling differentiable. Instead of sampling $z \sim \mathcal{N}(\mu,\sigma^2)$ directly, one samples $\epsilon \sim \mathcal{N}(0,1)$ and computes $z = \mu + \sigma \odot \epsilon$, allowing gradients to propagate through $\mu$ and $\sigma$.  

A wide variety of VAE architectures have been proposed in the literature, often tailored to exploit structure or symmetries in the input data. In our case, the goal is to learn multivariate distributions of six or seven variables. We employ an encoder consisting of two fully connected layers, the first with 256 hidden units and the second with 64 units, followed by two separate output layers producing $\mu$ and $\sigma$, each of dimension 16. The decoder mirrors the encoder architecture, mapping latent vectors back to the original data space.  

\subsubsection{Generative Adversarial Network (GAN)}\label{sec:GANS}

\begin{figure}[h!]
    \centering
    \includegraphics[width=0.7\linewidth]{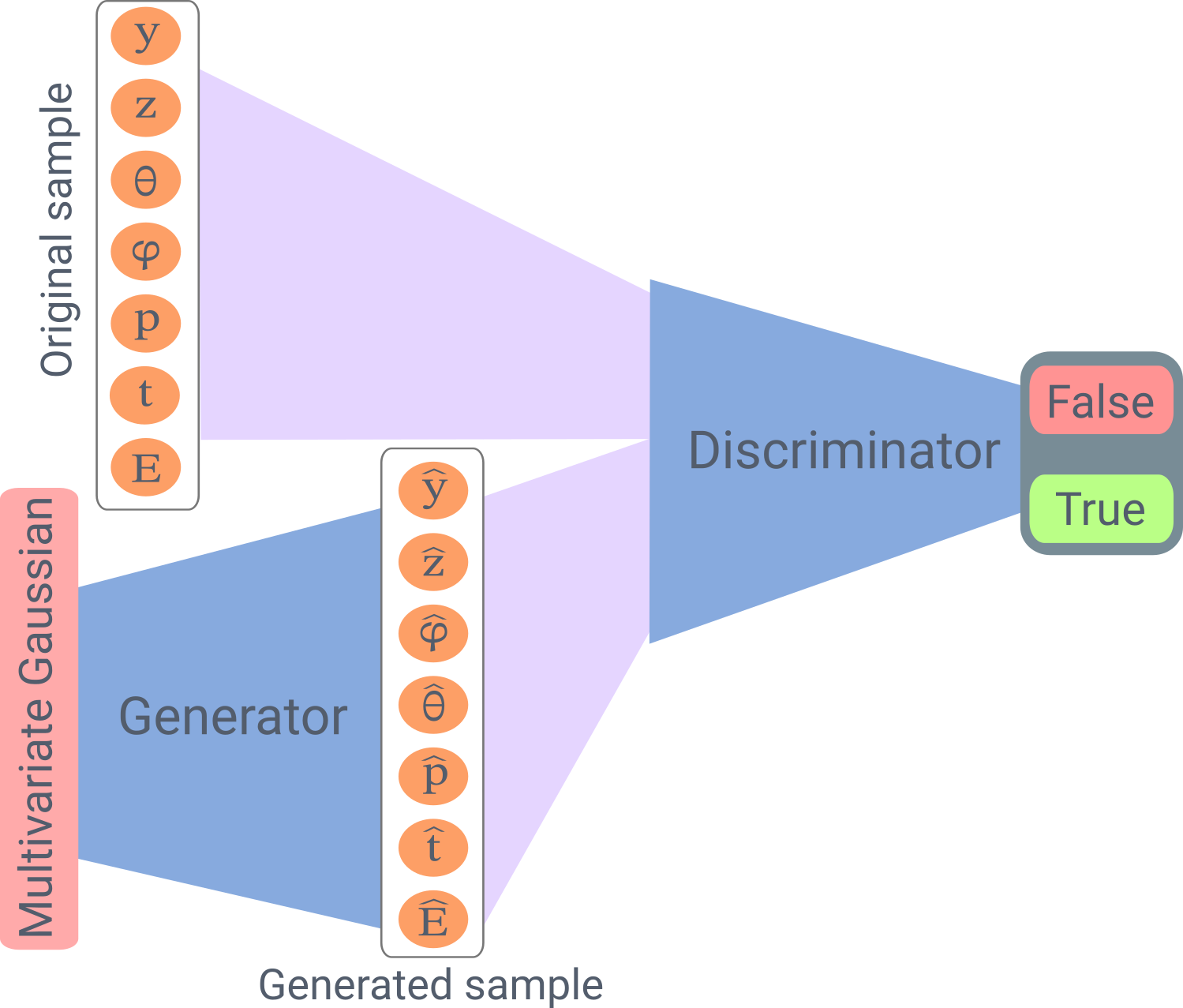}
    \caption{Conceptual schematic of a GAN model. The generator transforms a multivariate Gaussian random vector into an estimated sample, which then is compared to and original sample from the dataset by the discriminator in the attempt to discriminate them. If it cannot, the generator model has learned well the task, and can be used to generate new data that follow the original distribution.
    }
    \label{fig:schematic_GAN}
\end{figure}

The concept of a GAN has been proposed by Goodfellow et al. \cite{goodfellow2014}. They propose two networks that compete against each other in a game-theoretic scenario. One network is called the Generator and its objective is to generate fictitious data that resembles the original data, i.e. that belongs to the same data distribution as the dataset on which it was trained. The second network, the Discriminator, tries to classify the data generated by the Generator as belonging or not to the original dataset, i.e. it classifies as True if it cannot distinguish it from the original dataset and False if it does not belong to the original dataset distribution. A schematic of the architecture of a GAN is depicted in Fig. \ref{fig:schematic_GAN}, where the fictitious data is generated by means of sampling a multivariate Gaussian and then modified by passing through the generator to resemble an original sample. The Discriminator receives both generated and real samples. The loss function of a GAN is what gives it the adversarial nature: The discriminator is trained to correctly classify real and synthetic samples, typically by minimizing a binary cross-entropy loss, while the generator is trained to produce samples that the discriminator cannot distinguish from real (classifies as real). The total loss is computed by evaluating the output of the discriminator on generated samples, thereby quantifying how confidently the discriminator can classify them as synthetic. The generator's weights are updated to reduce this confidence, effectively encouraging the generation process to be more realistic with respect to the original distribution.  

By applying a learning scheme that alternately adjusts Generator and Discrimininator weights, it is possible to train a Generator that learns to generate data that belongs to the training dataset distribution. While GANs have shown remarkable results on image data (see e.g.~Ref.~\cite{karras2019}), GANs are faced with issues like mode collapse, where only a small subset of the training data is generated, and therefore commonly attributed to be difficult to train.

The Discriminator network used here is a convolutional neural network that grabs a normalized neutron phase-space variable vector and outputs a unique value that is 0 or 1. The process is done by means of 3 convolution layers. The Generator network is a deconvolutional neural network that transforms a 64 dimensional multivariate Gaussian input by means of three 1D transposed convolutional layers, into the normalized neutron phase-space vector that is then evaluated by the Discriminator. As with the NFs, A Sigmoid layer is added as the last layer to guarantee that the generated values lie in the [0,1] interval.

Although convolutional neural networks are most commonly associated with spatially structured data, the use of one-dimensional convolutions in this context does not rely on assumptions of spatial locality or translation invariance in the neutron phase-space variables. Instead, the one-dimensional convolutional layers are employed as parameter-efficient linear operators that enable structured mixing of input features across neighboring dimensions. That is, they act as shared-weight projections over the ordered phase-space vector, providing an efficient mechanism to capture correlations between subsets of variables while significantly reducing the number of trainable parameters compared to fully connected layers of comparable expressive power. This architectural choice allows for compact models with reduced risk of over-fitting, particularly advantageous in moderate to small size of training datasets. It is important to note that the ordering of variables is fixed and consistent across training and inference, and no physical notion of spatial locality is assumed or imposed.

\subsubsection{Diffusion Model (DM)}\label{sec:DM}
\begin{figure}[h!]
    \centering
    \includegraphics[width=0.5\linewidth]{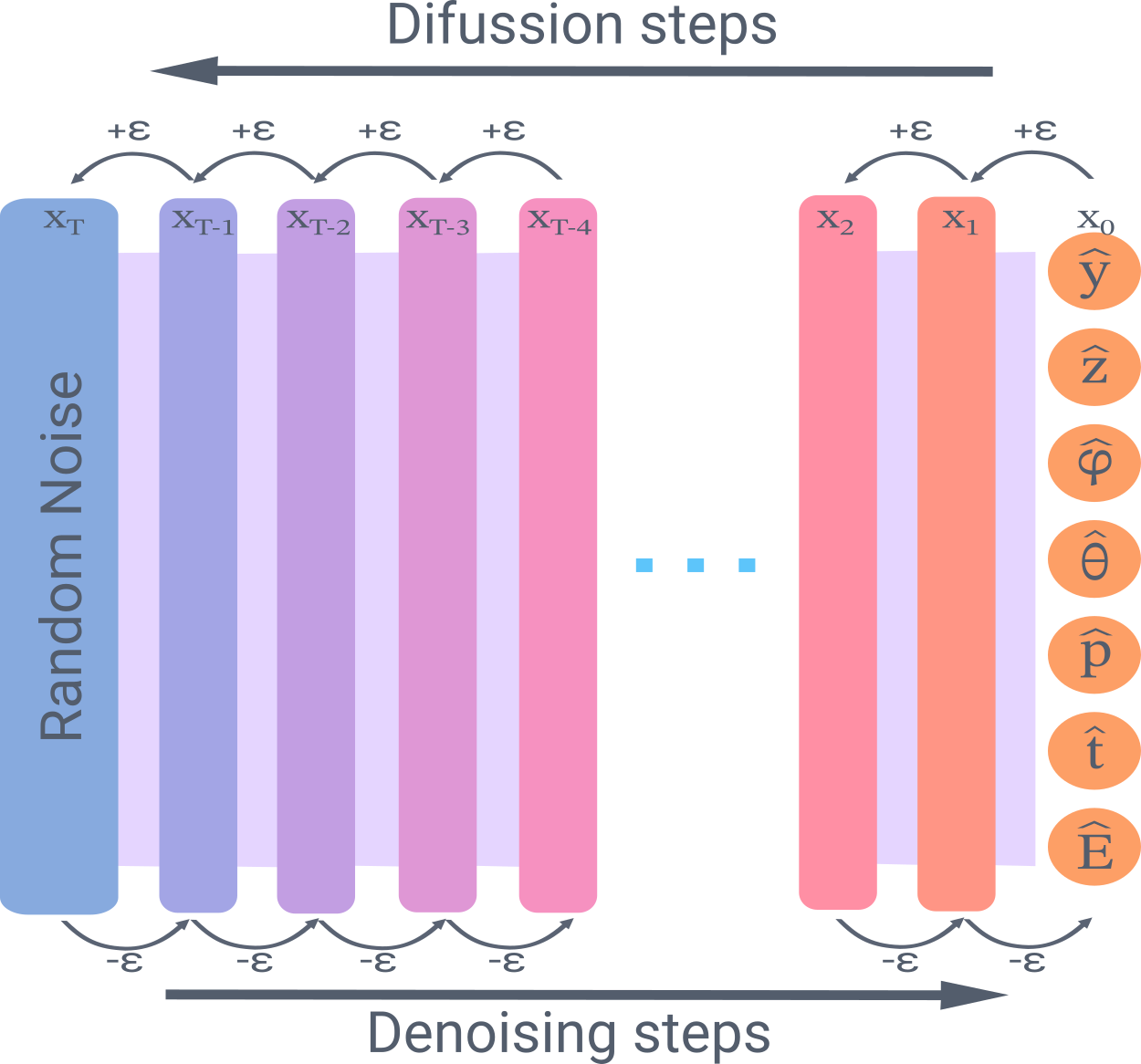}
    \caption{Conceptual schematic of a DM model. Starting from random noise, the denoising steps remove noise gradually in $T$ steps, until an estimate of the phase-space variables of a neutron is obtained.
    }
    \label{fig:schematic_DM}
\end{figure}

These models were introduced as a novel approach to generate high-quality data by simulating the gradual process of adding and then reversing noise. The core idea, as detailed by Sohl-Dickstein et al. \cite{sohl2015deep}, involves a forward process and a reverse process. In the forward process, noise is incrementally added to the data until the original data distribution is transformed into a Gaussian noise distribution. The reverse process, which is the generative aspect of the model, involves learning to reverse this noise addition to reconstruct the original data from the noise. A schematic of this process is depicted in Fig. \ref{fig:schematic_DM}. The model operates by training a neural network to parameterize the reverse diffusion process. This network predicts the noise that was added at each step of the forward process, effectively learning to denoise the data. As the training progresses, the model becomes better at reconstructing the original data from its noised states, thereby learning the data distribution. The training objective in DMs is typically framed as a variational inference problem, where the goal is to maximize the likelihood of the data given the model. This is achieved by minimizing a loss function that measures the difference between the original data and the data reconstructed by the reverse process at each step of the diffusion. It is noteworthy to mention that the denoising process might be very time-consuming because of the amount of steps of sequential removal of noise in generating new samples.

In this work, we developed a DM that consists of two primary components: a noise predictor network and a reverse diffusion process. The noise predictor network is a neural network designed to estimate the noise added during the forward diffusion process. It comprises of three linear layers, each with 128 hidden units, and utilizes Rectified Linear Units (ReLU, $f(x) = \max(0, x)$) as activation functions. The input to this network is a concatenation of the perturbed data and a time-encoded vector, which provides temporal context to the model.
The reverse process, built upon the foundation of the forward process, aims to reconstruct the original data from its noised version by iteratively denoising across time steps. This process uses the noise predictions from the noise predictor network to gradually remove the added noise. Specifically, the reverse process leverages a series of calculated noise standard deviations and mean adjustments based on the alphas (scaling factors for noise reduction) and betas (noise levels) defined in the forward process. These parameters guide the reverse diffusion to effectively denoise the data, aiming to restore the original data distribution.
Additionally, the reverse process includes a method to ensure that the generated samples remain within the unit interval [0,1]. This is achieved through a mirroring technique that folds and reflects values outside the unit interval back into it. A total of $T=200$ time steps was used to train the model. 

\subsection{Implementation}\label{sec:implementation}

The framework chosen for the training of the AI models was \texttt{PyTorch} \cite{pytorch}, as it is popular and very well documented, and it also provides \texttt{libtorch}, which is the C++ API that acts as a bridge to Vitess, the MC software written in C that is maintained at Forschungszentrum Jülich (FZJ). All models were trained for 1000 epochs, using 1 node of the JUWELS booster cluster \cite{juwels2019} of the Jülich Supercomputing Centre (JSC) at FZJ, Germany. These nodes have four A100 Graphical Processing units (GPUs), and the models were trained on PyTorch using a Data Distributed Parallel (DDP) approach for the parallelization on all GPUs, with a batch size of 4096. This resulted in an average training time of 8 minutes for all models trained in both datasets. All parameters of the models described in the previous subsections were found by using the \texttt{Optuna} library for optimal hyper-parameter search \cite{akiba2019optuna}.

 In the case of the benchmark dataset, after training all PGMs in Python, the models are saved as a \texttt{torch.jit.script} models which can then be loaded into a C++ module in Vitess called \texttt{source\_AI} that we developed specifically for the purpose of sampling on the flight the trained probabilistic models. This module decodes and inverse transforms the sampled values, and then feeds the generated neutrons to the next modules in Vitess, i.e. it acts as a source module for the Vitess package. 
It is noteworthy that components for McStas \cite{mcstas} or other neutron propagation software can be developed, but having a python interface is an advantage, allowing to sample directly within a jupyter notebook and then passing the generated neutrons to the simulation. This last step can be easily done for example by creating new MCPL files of the generated neutrons and using existing components to load MCPL files like \texttt{read\_in} in Vitess or \texttt{MCPL\_input} in McStas.

\subsection{Metric for comparing mutlivariate distributions}\label{sec:MMD}

It is imperative to have an accurate metric to compare and assess the quality of the multivariate probability distribution learned by the different models. A useful statistical test to determine if two samples are drawn from the same multivariate distribution is the Maximum Mean Discrepancy Test (MMD) \cite{gretton2012kernel}. The test functions by measuring the distance between the mean embeddings of features in a reproducing kernel Hilbert space (RKHS). 

Suppose we want to know if two samples $X$ and $Y$ were sampled from the same distribution. Let's suppose $X\sim P$ and $Y\sim Q$. We want a measure to inform us how similar is $P$ to $Q$. MMD computes the norm of the difference between the means of the distributions $\mu_P$ and $\mu_Q$ in the RKHS defined by kernel $K$. Let $\langle,\rangle$ denote the inner product of the Hilbert space, then

\begin{equation}
    ||\mu_P-\mu_Q||^2=\langle \mu_P-\mu_Q, \mu_P-\mu_Q \rangle = \langle \mu_P, \mu_P \rangle + \langle \mu_Q, \mu_Q \rangle - 2 \langle \mu_P, \mu_Q \rangle.
\end{equation}
Taking into account the reproducing property of the RKHS, it is possible to see that $\langle \mu_P , \mu_Q \rangle = \mathbb{E}_{X\sim P, Y\sim Q} \left[ K(X, Y) \right]$. Then, we arrive to the equation employed to computationally calculate the MMD from samples $X$ and $Y$,
\begin{equation}\label{eq:mmd}
    ||\mu_P-\mu_Q||^2 = \mathbb{E}_{X,X'\sim P} \left[ K(X, X') \right] + \mathbb{E}_{Y,Y'\sim Q} \left[ K(Y,Y') \right] -2\mathbb{E}_{X\sim P, Y\sim Q} \left[ K(X, Y) \right].
\end{equation}
Due to the correlation term in eq. \ref{eq:mmd}, means from  samples of the same distribution should lie near in the embedding space (i.e. distance near zero), and the distance should increase as the difference between the distributions increases.
The kernel functions can be interpreted to map the samples into a higher-dimensional space where the difference between the distributions become more evident. In this work, we used the Radial Basis Functions (RBFs) as kernel functions, defined as $K(x,y) = \exp(-||x-y||^2)$. 

We employ the MMD score to compare all generated samples to the original sample of the respective MCPL file. To have a notion of distance in this space, the MMD value of two samples from the original MCPL file is also calculated as a ``near" distance value, and the MMD value of a sample from a uniform distribution in all phase-space variables against a sample of the original MCPL file is used as a ``far away" reference value. This is understood as the less informative distribution estimator.

\section{Results}

\subsection{Model comparison}

Following the training of both the TDR and benchmark datasets as outlined in the previous section, the models under consideration exhibited a notable capacity to learn a reasonable representation of the underlying probability distributions in both datasets. The Maximum Mean Discrepancy (MMD) value between a sample from the MCPL file and a sample from all models, as well as with a sample from itself (MCPL sample) and a fictitious sample from a uniform distribution in all variables (Uniform sample) are presented in Table \ref{tab:mmd_values}. 
\begin{table}[ht]
\centering
\caption{MMD Values for Different Models on TDR and Benchmark Datasets. These values are calculated between a sample of 20000 neutrons from the MCPL file (original distribution sample) and a sample of the same size from the proposed models. Sampling was performed 5 times and mean value and standard deviation of the MMD values on each iteration are shown. }
\label{tab:mmd_values}
\begin{tabular}{@{}lcc@{}}
\hline
Sampling from & MMD - TDR [$\times 10^{-4}$] & MMD - Benchmark [$\times 10^{-4}$] \\ 
\hline
\hline
DM          & $1.33 \pm 0.03$ & $1.95 \pm 0.05$ \\
VAE & $1.22 \pm 0.02$ & $1.70 \pm 0.03$ \\
NF-MAF        & $1.21 \pm 0.02$ & $\mathbf{1.17 \pm 0.03}$ \\
NF-Coupling   & $\mathbf{1.19 \pm 0.02}$ & $1.20 \pm 0.04$ \\
GAN & $1.28 \pm 0.03$ & $1.81 \pm 0.05$ \\
\hline
KDSource                      & -                       & $1.58 \pm 0.02$ \\
MCPL                  & $0.96 \pm 0.01$ & $0.83 \pm 0.03$ \\
uninformative Uniform                & $5.86 \pm 0.02$ & $14.93 \pm 0.05$ \\
\hline
\end{tabular}
\end{table}
All MMD values were calculated with samples of 20.000 neutrons and the random samples were iterated 5 times to calculate the standard deviation. In addition, the KDSource approach from \cite{schmidt_2025} was utilized for comparison. However, this value is only available in the Benchmark dataset, as KDSource has not yet the capability to estimate the particle's weight distribution, which is present in the TDR dataset. A comparison of the MMD values indicates that NFs have demonstrated superiority in learning the multivariate distribution, although all models are capable of estimating the underlying distribution. We do observe that, in terms of MMD, the performance of each model depends on the dataset features. While it is not possible to formulate a model that can accomodate all potential datasets, NFs both with coupling flows or with masked autoregressive flows have demonstrated the greatest degree of flexibility. It appears that learning multimodal distributions is easier for MAFs, but having uniform distributions on x, and y directions in a square (sharp features) seems more challenging to learn for MAFs than for coupling flows, making them in the end have very similar performances.

\begin{table}[ht]
\centering
\caption{Average sampling time of 20,000 Neutrons for each model on both datasets. Average and standard deviation values are calculated in 5 repetitions. }
\label{tab:sampling_time}
\begin{tabular}{@{}lcc@{}}
\hline
Model & TDR Dataset (s) & Benchmark Dataset (s) \\ 
\hline
\hline
DM & $1.972 \pm 0.040$ & $1.875 \pm 0.050$ \\
VAE & $0.0165 \pm 0.0006$ & $0.0161 \pm 0.0028$ \\
NF-MAF & $0.3222 \pm 0.0092$ & $0.2942 \pm 0.0283$ \\
NF-Coupling & $0.0370 \pm 0.0008$ & $0.0158 \pm 0.0002$ \\ 
GAN & $0.0401 \pm 0.0012$ & $0.0231 \pm 0.0031$ \\
KDSource & - & $0.0122 \pm 00008$ \\
\hline
\end{tabular}
\end{table}

The results of the MMD values in the Benchmark dataset demonstrate a better performance of NF models over the state-of-the-art approach, KDSource. The proposed model offers three main advantages. First, the NF has learned the distribution, eliminating the need of the initial dataset (i.e. it is independent from the original MCPL file, which is not the case in KDSource). Furthermore, no special consideration needs to be taken into account to estimate the multivariate phase-space (black-box implementation), while on KDSource one must tune the importance weights for variables. The extension of NFs to the entire phase-space of the neutron can be achieved without any additional consideration, apart from increasing the dimension of the inputs, while KDSource is implemented only for a subset of the phase-space variables and therefore can be problematic for some scenarios (like in the TDR dataset with the variance reduction weight estimation, or in any case with polarization). Nevertheless, both estimations are noteworthy approaches and have potential to be used in a variety of scenarios. KDSource has the advantage of faster sampling.  Furthermore, KDSource is more adept at handling cases of very low statistics (low number of neutrons, where the ML models may fail to learn distributions and over-fit).

\subsection{TDR dataset}

\begin{figure}[h!]
    \centering
    \includegraphics[width=1\linewidth]{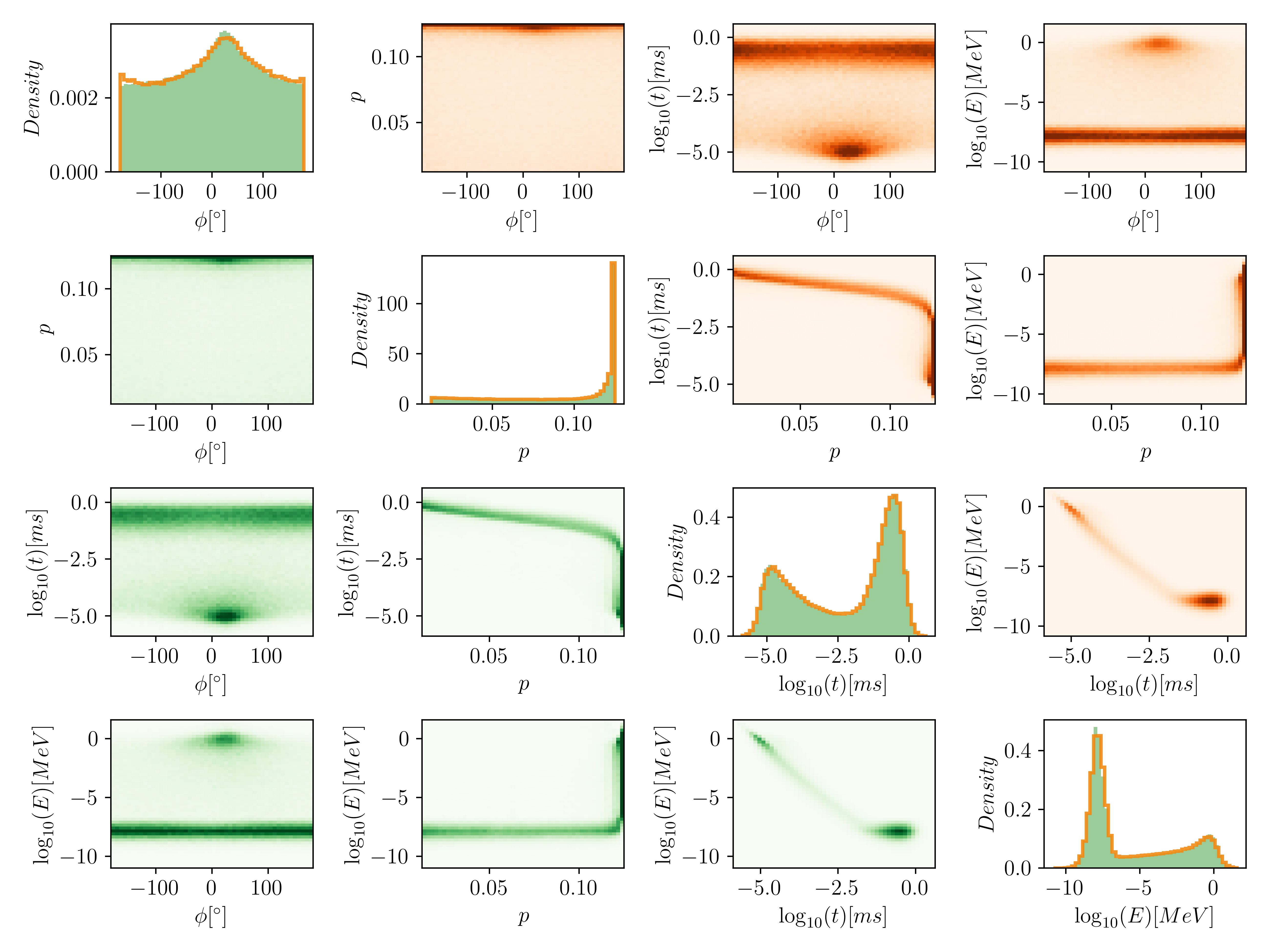}
    \caption{Histograms for azimuthal angle, weight, time and energy variables for TDR dataset. 1D Histograms in the diagonal show the marginal distributions for a sample from the original data (green) and for the AI model (orange). 2D histograms below the diagonal correspond to the data, and those above the diagonal to the NF AI model.}
    \label{fig:generated_distribution_nflow_tdr2}
\end{figure}

Following the results from the previous section, the NF model with coupling layers was used to generate a sample of the same size as the original MCPL file of the TDR dataset (676.558 neutrons). These neutrons were then used to construct the histograms shown in Fig.  \ref{fig:generated_distribution_nflow_tdr2}. The marginal distributions of some of the variables of the TDR dataset can be seen in the diagional plots in green and the histograms generated with a sample from the the NF model can be seen in orange steps, practically outlining that of the data. This emphasizes the agreement in the quantitative comparison result of these two samples by means of MMD shown in Table \ref{tab:mmd_values}. The variables shown in the pairplot are the ones with highest correlations. For a complete comparison of all 7 estimated variables, see the suplementary material figures. 

 Fig. \ref{fig:generated_distribution_nflow_tdr2} also shows the 2D histograms of the original data on the plots below the diagonal (colored in green), and the 2D histograms of the generated data above the diagonal (in orange) for an easy comparison. Most importantly, from a comparison of the plots below the diagonal with those above the diagonal in Fig. \ref{fig:generated_distribution_nflow_tdr2} we can confirm that the correlations between variables are preserved when using the NF model to generate new neutrons. This can be seen by inspecting the 2D histograms of time and energy, weight and energy, or angular distribution and position variables. Small features of this specific dataset are also learned, for example the fact that fast neutrons are not isotropic in $\varphi$, but are rather centered around $\varphi=28^{\circ}$, which is a consequence of the fact that the fast neutrons are emitted from the target which is positioned below the extraction layer and shows an angle of $28^{\circ}$ in the vertical plane with the surface where the particles are recorded.
 
\subsection{Benchmark dataset}

\begin{figure}[h!]
    \centering
    \includegraphics[width=1\linewidth]{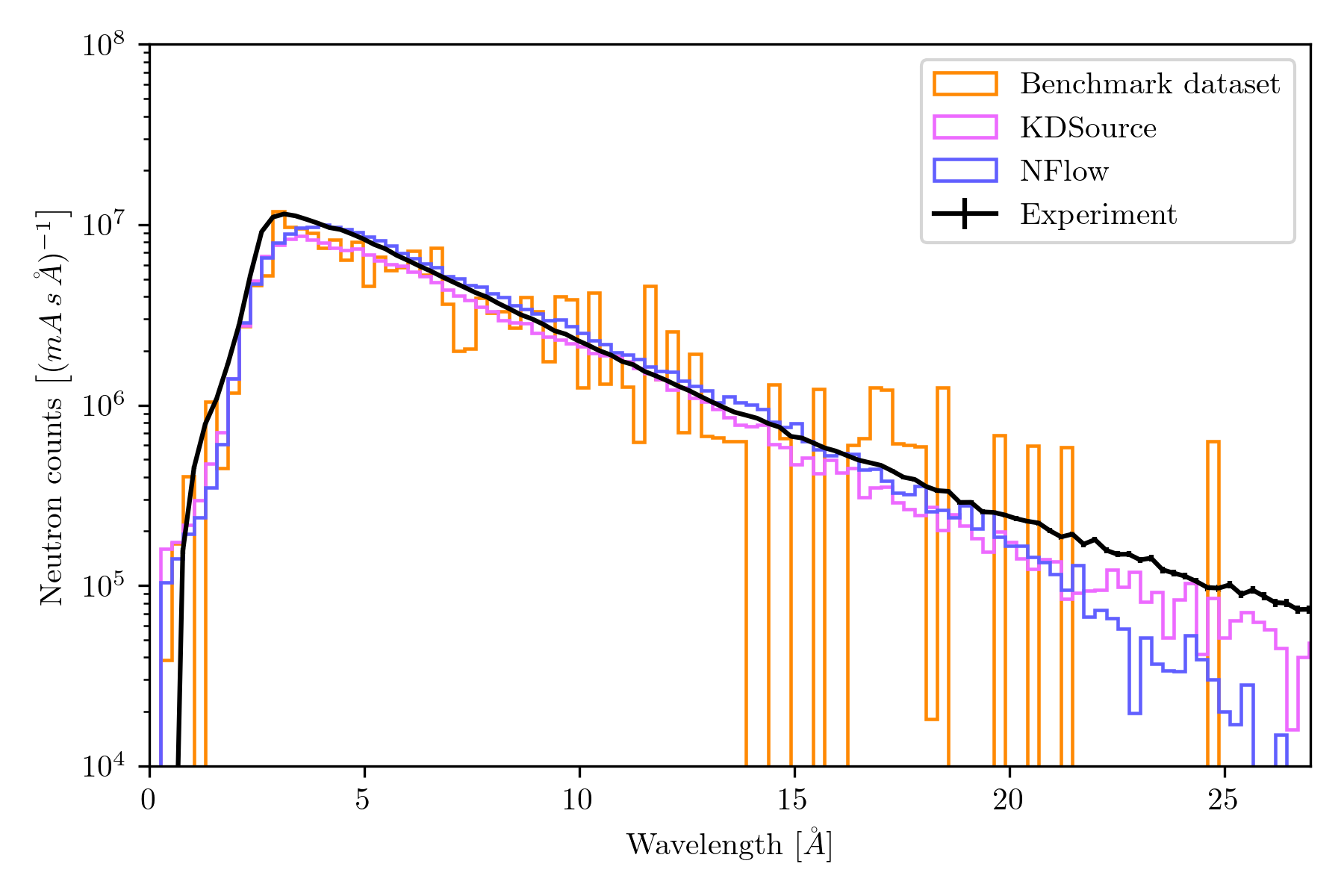}
    \caption{Comparison of simulated and measured spectra of the J\"ulich platform experiments. the results of simulations using different simulation source estimates are shown.}
    \label{fig:comparison_experimental}
\end{figure}

\begin{figure}[h!]
    \centering
    \includegraphics[width=1\linewidth]{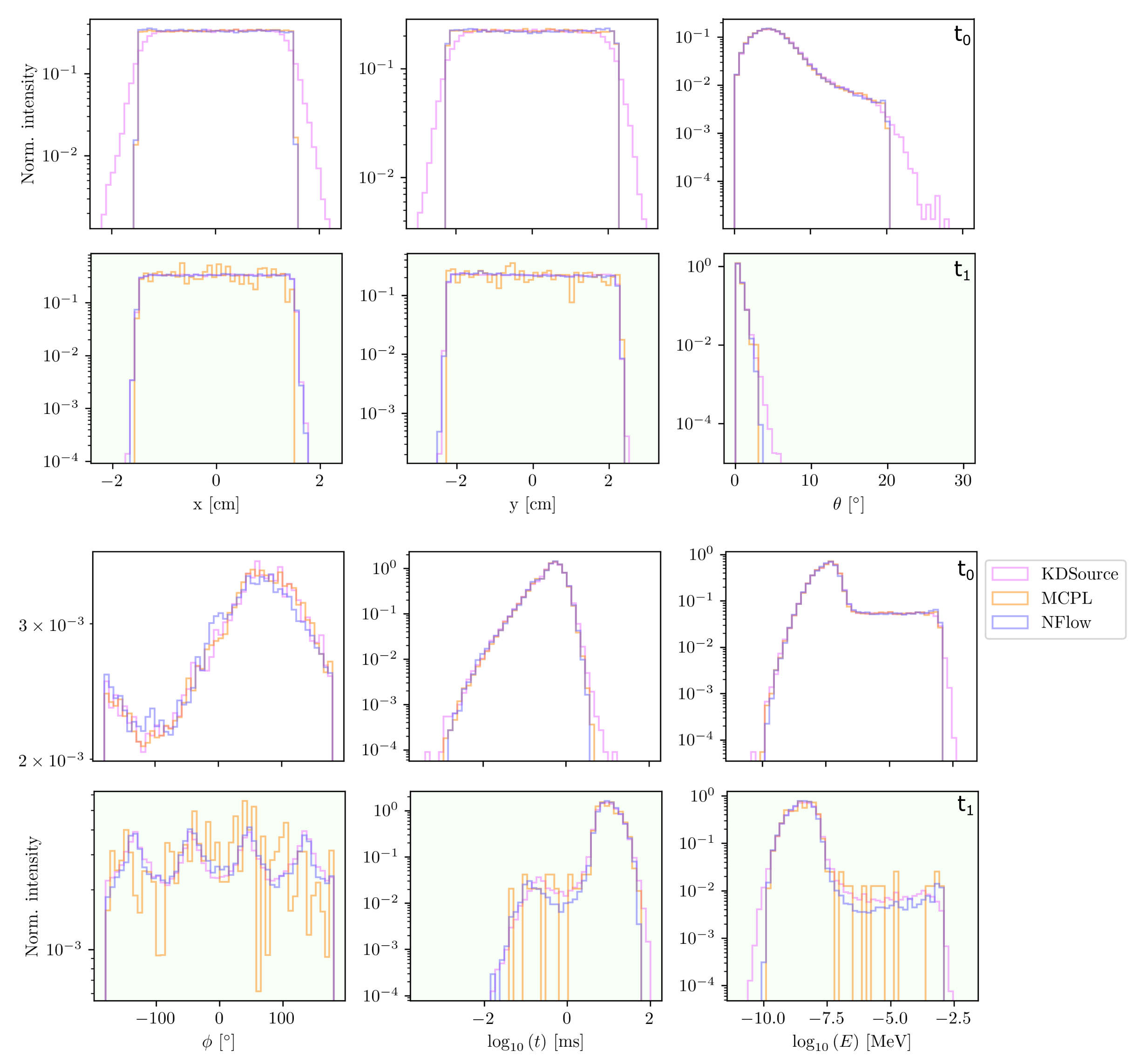}
    \caption{Comparison of the results of the Monte Carlo simulations on VITESS, using the MCPL as input file (orange), the KDSource (pink), and the NF model (blue). Marginal distributions of the estimated phase-space variables are shown at two different points of the simulation: at the beginning just after the source module ($t_0$, white background) and at the end of the simulation just before the detector ($t_1$, green background). on all $t_0$ plots, MCPL and NF overlap with minimal discrepancies.}
    \label{fig:marginals}
\end{figure}

One of the results from the experiments conducted at the J\"ulich platform was the time-of-flight spectrum measured by a $^3$He detector, which was then converted to a wavelength spectrum following the procedure described in \cite{schmidt_2025}. A comparison of the spectrum at the detector for all different approaches with the experimentally measured spectrum  is presented in Fig. \ref{fig:comparison_experimental}. It is possible to see that both KDSource and the NF approach effectively reproduce the primary features of the observed wavelength spectrum. Furthermore, it also demonstrates that the repetition of the MCPL file to reach $1\times10^8$ neutrons leads to the generation of hot spots. A preliminary analysis of the wavelength spectrum suggests that KDSource exhibits superior performance for large wavelengths. However, a thorough examination of the Benchmark dataset result reveals that the MCPL file utilized does not contain any neutrons with wavelength above 25$\AA$. This observation indicates that this region of wavelengths is likely to be an extrapolation. Given that the NF model is constrained to learn from observed data and that the domains for generated neutrons are forced to the maximum and minimum values encountered in the training dataset, it is not possible for the model to learn features outside of these bounds. In contrast, with KDSource, it is possible to go beyond the data boundaries depending on the selected kernel and bandwidth. This does not imply that NFs are incapable of extrapolation. To allow the NF to learn outside of the boundaries, it is sufficient to remove the Sigmoid layer from the architecture that restricts the values to the interval [0,1] (prior to undoing the MinMax normalization). This feature of KDSource can be seen emphasized in all the marginal distributions of the estimated phase-space variables in Fig. \ref{fig:marginals}. The upper row of this figure displays the monitor readings of the estimated phase-space variables at the beginning of the simulation, while the lower row presents the same readings just before they reach the detector. It is evident that both KDSource and NF provide an accurate estimation of the MCPL file distribution (orange curve) by examining the overlap of the marginal distributions on the upper row. The NF and Benchmark Dataset histograms exhibit almost perfect overlap in all variables, while KDSource has edge discrepancies, which are purposely emphasized in the log-scale nature of these plots. The disparities between KDSource and the original dataset are most pronounced in the $x$ and $y$ variables, though they are also evident in all marginal distributions. The application of a kernel density estimation based on a Gaussian kernel revealed the potential presence of neutrons outside the designated guide, as well as faster neutrons that were not present in the dataset. This phenomenon manifests a favorable outcome in the context of cold neutrons. The presence of colder neutrons, which were not previously present in the MCPL file, becomes evident. These neutrons are capable of traversing the neutron guide and reaching the detector, as illustrated in the final column plots of Figure \ref{fig:marginals}. In contrast, the implementation of a NF did not yield any neutrons outside the anticipated phase-space domain which correctly reproduces the experiment in the case of the $x$ and $y$ variables. With respect to divergence, KDSource also overestimates the domain; however, these effects are negligible and do not significantly impact the simulation output.

The improved performance observed for the NF model should not be attributed solely to preprocessing, but rather to the ability of the architecture to incorporate hard physical constraints in a principled manner. By construction, NFs define an explicit, invertible mapping between a simple base distribution and a target distribution. This property allows domain constraints, such as bounded support enforced through normalization and sigmoid constraints, to be embedded directly into the generative mapping, ensuring that all generated samples lie within physically admissible regions of phase-space. While the expressive power required to learn the underlying probability distribution is determined by the model architecture itself, the inclusion of hard constraints can significantly simplify the learning task. In this sense, constraints act as inductive biases that guide model toward physically meaningful solutions. Although similar constraints could in principle be imposed on other generative models, their integration is less straightforward and often requires additional architectural components or post-processing steps. 

\section{Discussion}

The main difference between the proposed algorithms lies on the way they are trained to learn probability distributions. Each one of these models has advantages and disadvantages, which should be taken into account when selecting a model to reproduce a source distribution from an MCPL file.

The advantages of a VAE over the other models are that it can learn very complex distributions and is relatively fast to sample from. The disadvantages are that it may suffer mode collapse, which deviates from the original distribution and can have a tough time during training to avoid mode collapse. Also, they are known for generative more ``blurry" data, i.e. with some noise in it, therefore less accurate.

The advantage we see on NFs are the capability of generating very accurate data, i.e. they are excellent in learning the underlying probability distribution of the MCPL file. The problem with them is that they are usually slower to sample from, and might have difficulties in expressing very sharp features in the variable distributions due to the invertible nature of the transformations applied.

DMs have shown good potential for estimating the phase-space distribution, but are the hardest models to define in terms of architecture and they are the slowest to train and to sample from. For simple tasks such as the one presented in this work, they might be too complex.

Finally, although GANs are known for their high accuracy in the generated images, they can also suffer mode collapse, and have proven to be the hardest to train in the task stated in this work, achieving a reasonable estimation only after many iterations. This might be due to the compromise between both loss terms described in the \ref{sec:GANS} section.

From the authors perspective, NFs are the simplest and most parsimonious approach for non Machine learning experts when needing to implement a model. They have shown to be the most robust when training models from MCPL files, and have shown the lowest MMD, as well as lowest sampling times in our benchmark task.

All of these AI models allow the addition of physically informed loss terms to the total loss function to add constraints on the features learned from the dataset. This is nowadays called physics informed neural networks (PINNs) and is a subject of very high potential usage when it comes to source distribution estimation for more complex and constraint scenarios. A final comment is that, in all of these models, it is possible to study the latent space representation, to correlate different features of the generated neutrons with specific volumes of the latent space. In this sense, it is possible to gain interpretability in the latent space, and then perform conditional sampling as necessary. The complexity of the latent space varies on each model, and can be set to suit specific applications.

In this work, the same model architectures were applied across two different neutron source problems, and in both cases the selected architectures required no substantial retuning, beyond trivial adaptations related to the dimensionality of the phase-space representation (i.e., the number of sampled variables). No case required problem-specific redesign of the model structure or extensive hyperparameter optimization, and comparable performance was obtained across all investigated generative paradigms. This is an indicator of robustness that is essential for practical deployment in neutronics applications.

While generative machine learning models provide a powerful alternative to the direct reuse of particle lists or to repeating Monte Carlo source simulations, their adoption introduces a set of challenges and trade-offs that must be carefully considered. The workflow investigated in this work does not aim to universally replace repeated MC simulations, but rather to provide an efficient and flexible surrogate in scenarios where repeated high-statistics simulations are computationally prohibitive or time consuming.

One of the primary considerations concerns model selection and architectural complexity. As shown in section \ref{sec:methods}, generative ML models require choices regarding network architecture, latent dimensionality, loss functions, and training strategies. These choices can influence the fidelity of the learned distribution and may require domain expertise, as well as iterative tuning. Training stability also represents a non-negligible cost. Another important limitation is the risk of mode collapse or incomplete support coverage. Such failures may manifest as oversampling on high density regions or missing particles from rare event regions of the phase space volume. While techniques as regularization, architectural constraints, and alternative generative paradigms can mitigate these effects, they do not eliminate the need for careful validation. Therefore, a quantitative diagnostic and comparison against reference MC simulations and experimental benchmarks, as the one performed in this work, should be established as essential components of any generative source modeling workflow.

It should be noted that generative models are inherently data-driven and therefore limited by the quality of the training dataset. Any systematic bias, under-sampling, or modeling deficiency present in the original MC simulation will inevitably be learned by the model. However, unlike direct reuse of particle lists, generative models can partially alleviate statistical noise by learning smooth approximations of the underlying distribution, provided that the training data sufficiently covers the relevant phase-space.

The apparent modest size of the datasets considered in this work reflects a constraint common in neutron source simulations, where generating very large particle lists can be computationally expensive. Rather than representing a limitation, this setting highlights one of the key advantages of the proposed approach. The considered generative models are trained to capture the joint structure of the phase-space distribution rather than to memorize individual samples, enabling them to learn smooth, correlation-preserving representations even from relatively limited data, provided that the relevant regions of phase-space are adequately sampled. This feature becomes particularly interesting in practical simulation scenarios where oversampling is required. When sampling more particles than those available in the original MCPL file, direct reuse necessarily repeats identical neutrons, whereas generative models produce new realizations consistent with the learned distribution. Moreover, in studies involving multiple instrument configurations, sample environments, or repeated simulations under varying conditions, it is often undesirable for identical neutron histories to be reused across different runs. In such cases, generative source models enable controlled variability while preserving the statistical properties of the source, improving robustness and reducing correlations between simulation outcomes.

From a practical perspective, the proposed approach is most advantageous in use cases where a high-fidelity source description must be reused across multiple instrument configurations or optimization studies. In this scenario, the one-time cost of model training can be amortized over many subsequent simulations, leading to substantial overall computational savings. The training phase seems naturally aligned with  the responsibilities of neutron sources or facilities, who already perform the underlying MC simulations. Once trained, the resulting generative models are lightweight and can be distributed to end users, who can sample form them without any knowledge of model architecture or training procedures.

\section{Conclusions}

In this work, we estimate the multivariate neutron phase-space variables distribution of a neutron source simulation by means of various probabilistic generative models and compare them to assess their advantages and disadvantages. We show that this is possible, and we perform a thorough comparison between recent generative models and traditional alternatives to source estimation. 

The relative performance of the considered generative models depends on the specific characteristics of the source distribution and the intended application. Amongst the probabilistic generative models studied, NFs have shown to work the best for this task, overcoming all other probabilistic density estimation and traditional methods. In particular, NFs are specially well suited to scenarios where the target distribution can be well represented through a sequence of invertible transformations. This result highlights that the selection of a generative model should be guided by the structure of the phase-space distribution and by the computational constraints, as well as the requirements of the downstream simulation workflow.

While the present study focuses on two representative datasets, the results indicate that the proposed approach is robust in low-data regimes typical of neutron source simulations and scalable in practical usage scenarios where trained models are reused extensively.

The possibility of learning the likelihood function of our source simulation from an MCPL file opens up an immense landscape of improvements towards optimization of neutron scattering instruments, shielding, and source geometry for future neutron research facilities.      

\vspace{6pt} 

\suppdata{
We provide The following pairplots to observe the distributions of all variables trained on each dataset that was used in the manuscript. It is possible to see that the TDR dataset (Fig. \ref{fig:distributions_tdr}) has 7 variables, and the benchmark dataset (Fig. \ref{fig:distributions_benchmark} has 6. This is because the weight variable in the benchmark dataset was constant with a value of 1 for all neutrons, therefore there was no need in predicting it. The energy ranges differ substantially because the TDR dataset includes cold, thermal and fast neutrons, while the benchmark dataset only has the cold and thermal part of the spectrum. The angular and spatial distribution are completely different. 

 \begin{figure}[h!]
    \centering
    \includegraphics[width=1\linewidth]{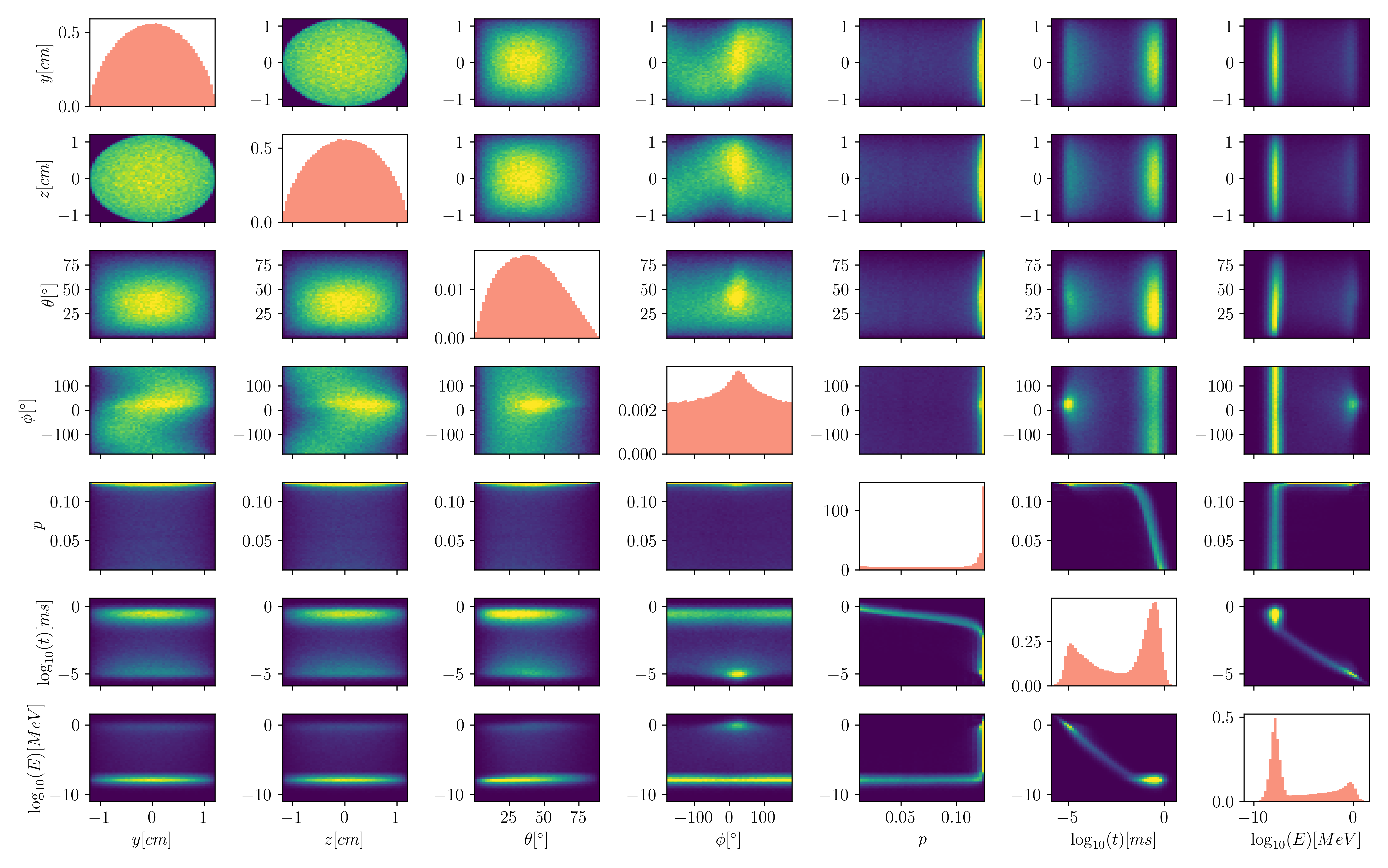}
    \caption{Pairplot of variables in the TDR dataset of the HBS project used to train PGMs. Individual histograms are shown on the diagonal and 2D histograms on the off-diagonals.}
    \label{fig:distributions_tdr}
\end{figure}

Fig. \ref{fig:distributions_tdr} attempts to give an idea of the complexity of the multivariate distribution of the variables of interest within the MCPL file. The marginal distributions are plotted by means of normalized histograms in the diagonal, while the 2-dimensional histograms corresponding to each pair of variables are shown on the off-diagonal plots. Observing this type of plot is of interest because it evidences that the considered variables are correlated in a complex way. There may exist higher-dimensional correlations, which are not visible in this plot, but that are very important to also have into consideration when modeling. Reproducing these correlations is a key requirement for the choice of the ML method.

 \begin{figure}[h!]
    \centering
    \includegraphics[width=1\linewidth]{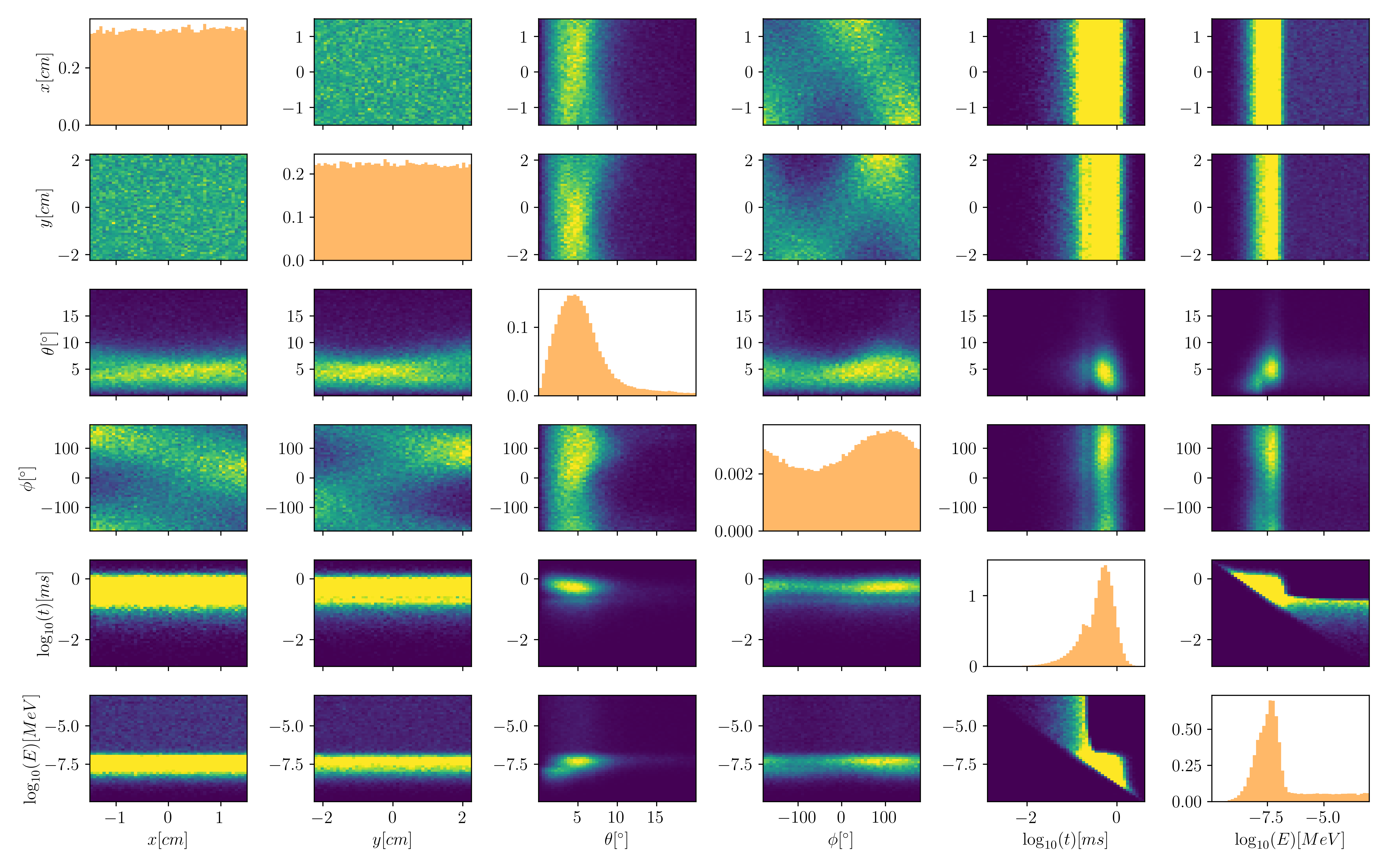}
    
    \caption{Pairplot of all transformed variables in the benchmark dataset of the Jülich platform simulations. Individual histograms are shown on the diagonal and 2D histograms on the off-diagonals. In this case, the weight was constant and not estimated by the PGM.}
    \label{fig:distributions_benchmark}
\end{figure}
}

\roles{Conceptualization, J.R., P.Z., and S.K.; methodology, J.R. and S.K.; software, J.R.; validation, J.R., and K.L.; formal analysis, J.R. and S.K.; investigation, J.R., P.Z.; resources, P.Z., and S.K.; data curation, N.S., J.L, and J.R.; writing---original draft preparation, J.R.; writing---review and editing, P.Z., S.K., and K.L.; visualization, J.R.; supervision, P.Z. and S.K.; project administration, S.K.; funding acquisition, P.Z., and S.K. All authors have read and agreed to the published version of the manuscript.}

\funding{This work was supported by Helmholtz AI Computing resources (HAICORE) of the Helmholtz Association's Initiative and Network Fund through Helmholtz AI. This work received funding from the European Union's Horizon 2020 research and innovation programme under the Marie Skłodowska-Curie grant agreement No 101034266}

\data{Data of the TDR and Benchmark datasets can be requested to the HBS project director.} 

\ack{The authors would like to acknowledge the High Brilliance Source project, in particular Ulrich Rücker for the measurements performed in the JULIC Neutron Platform during 2023.}

\bibliographystyle{iopart-num}
\bibliography{bibliography}

@inproceedings{forster2006mcnp,
  title={MCNP-a general Monte Carlo code for neutron and photon transport},
  author={Forster, RA and Godfrey, TNK},
  booktitle={Monte-Carlo Methods and Applications in Neutronics, Photonics and Statistical Physics: Proceedings of the Joint Los Alamos National Laboratory-Commissariat {\`a} l'Energie Atomique Meeting Held at Cadarache Castle, Provence, France April 22--26, 1985},
  pages={33--55},
  year={2006},
  organization={Springer}
}

@article{niita2006phits,
  title={PHITS—a particle and heavy ion transport code system},
  author={Niita, Koji and Sato, Tatsuhiko and Iwase, Hiroshi and Nose, Hiroyuki and Nakashima, Hiroshi and Sihver, Lembit},
  journal={Radiation measurements},
  volume={41},
  number={9-10},
  pages={1080--1090},
  year={2006},
  publisher={Elsevier}
}

@article{romano2015openmc,
  title={OpenMC: A state-of-the-art Monte Carlo code for research and development},
  author={Romano, Paul K and Horelik, Nicholas E and Herman, Bryan R and Nelson, Adam G and Forget, Benoit and Smith, Kord},
  journal={Annals of Nuclear Energy},
  volume={82},
  pages={90--97},
  year={2015},
  publisher={Elsevier}
}

@article{kittelmann2017monte,
  title={Monte Carlo particle lists: MCPL},
  author={Kittelmann, Thomas and Klinkby, Esben and Knudsen, Erik B and Willendrup, Peter and Cai, Xiao-Xiao and Kanaki, Kalliopi},
  journal={Computer Physics Communications},
  volume={218},
  pages={17--42},
  year={2017},
  publisher={Elsevier}
}

@article{willendrup2020mcstas,
  title={McStas (i): Introduction, use, and basic principles for ray-tracing simulations},
  author={Willendrup, Peter Kj{\ae}r and Lefmann, Kim},
  journal={Journal of Neutron Research},
  volume={22},
  number={1},
  pages={1--16},
  year={2020},
  publisher={IOS Press}
}

@article{mcstas,
author = {Willendrup, Peter Kjær and Lefmann, Kim},
title = {McStas (i): Introduction, use, and basic principles for ray-tracing simulations},
journal = {Journal of Neutron Research},
volume = {22},
number = {1},
pages = {1-16},
year = {2020},
doi = {https://doi.org/10.3233/JNR-190108},
URL = { 
        https://content.iospress.com/articles/journal-of-neutron-research/jnr190108
}
}

@article{mc-neutron1,
author = {G. Zsigmond and K. Lieutenant and F. Mezei},
title = {Monte Carlo simulations of neutron scattering instruments by VITESS: Virtual instrumentation tool for ESS},
journal = {Neutron News},
volume = {13},
number = {4},
pages = {11-14},
year = {2002},
publisher = {Taylor & Francis},
doi = {https://doi.org/10.1080/10448630208218488},
}

@article{lin2016mcvine,
  title={MCViNE--An object oriented Monte Carlo neutron ray tracing simulation package},
  author={Lin, Jiao YY and Smith, Hillary L and Granroth, Garrett E and Abernathy, Douglas L and Lumsden, Mark D and Winn, Barry and Aczel, Adam A and Aivazis, Michael and Fultz, Brent},
  journal={Nuclear Instruments and Methods in Physics Research Section A: Accelerators, Spectrometers, Detectors and Associated Equipment},
  volume={810},
  pages={86--99},
  year={2016},
  publisher={Elsevier}
}

@article{mc-neutron4,
title = {Prompt: Probability-conserved cross section biasing Monte Carlo particle transport system},
journal = {Computer Physics Communications},
volume = {295},
pages = {109004},
year = {2024},
issn = {0010-4655},
doi = {https://doi.org/10.1016/j.cpc.2023.109004},
url = {https://www.sciencedirect.com/science/article/pii/S0010465523003491},
author = {Zi-Yi Pan and Ni Yang and Ming Tang and Peixun Shen and Xiao-Xiao Cai}
}

@article{KDSOURCE,
  title={KDSource, a tool for the generation of Monte Carlo particle sources using kernel density estimation},
  author={Schmidt, Norberto Sebasti{\'a}n and Abbate, Osiris Inti and Prieto, Zoe Micaela and Robledo, Jos{\'e} Ignacio and Dami{\'a}n, JI M{\'a}rquez and Marquez, Ariel An{\'\i}bal and Dawidowski, Javier},
  journal={Annals of Nuclear Energy},
  volume={177},
  pages={109309},
  year={2022},
  publisher={Elsevier}
}

@article{forget2022normalizing,
  title={Normalizing flows for thermal scattering sampling},
  author={Forget, Benoit and Alhajri, Abdulla},
  journal={Annals of Nuclear Energy},
  volume={170},
  pages={108974},
  year={2022},
  publisher={Elsevier}
}

@article{berahmand2024autoencoders,
  title={Autoencoders and their applications in machine learning: a survey},
  author={Berahmand, Kamal and Daneshfar, Fatemeh and Salehi, Elaheh Sadat and Li, Yuefeng and Xu, Yue},
  journal={Artificial Intelligence Review},
  volume={57},
  number={2},
  pages={28},
  year={2024},
  publisher={Springer}
}

@article{singh2021overview,
  title={An overview of variational autoencoders for source separation, finance, and bio-signal applications},
  author={Singh, Aman and Ogunfunmi, Tokunbo},
  journal={Entropy},
  volume={24},
  number={1},
  pages={55},
  year={2021},
  publisher={MDPI}
}

@article{pytorch,
  title={Pytorch: An imperative style, high-performance deep learning library},
  author={Paszke, Adam and Gross, Sam and Massa, Francisco and Lerer, Adam and Bradbury, James and Chanan, Gregory and Killeen, Trevor and Lin, Zeming and Gimelshein, Natalia and Antiga, Luca and others},
  journal={Advances in neural information processing systems},
  volume={32},
  year={2019}
}

@inproceedings{akiba2019optuna,
  title={Optuna: A next-generation hyperparameter optimization framework},
  author={Akiba, Takuya and Sano, Shotaro and Yanase, Toshihiko and Ohta, Takeru and Koyama, Masanori},
  booktitle={Proceedings of the 25th ACM SIGKDD international conference on knowledge discovery \& data mining},
  pages={2623--2631},
  year={2019}
}

@article{Sarrut_2019,
doi = {10.1088/1361-6560/ab3fc1},
url = {https://dx.doi.org/10.1088/1361-6560/ab3fc1},
year = {2019},
month = {oct},
publisher = {IOP Publishing},
volume = {64},
number = {21},
pages = {215004},
author = {D Sarrut and N Krah and J M Létang},
title = {Generative adversarial networks (GAN) for compact beam source modelling in Monte Carlo simulations},
journal = {Physics in Medicine \& Biology}
}

@article{ refId0,
	author = {{Manring, C. A.} and {Hawari, A. I.}},
	title = {Development of neural thermal scattering (NeTS) modules for reactor multi-physics simulations},
	DOI= "10.1051/epjconf/202124720004",
	url= "https://doi.org/10.1051/epjconf/202124720004",
	journal = {EPJ Web Conf.},
	year = 2021,
	volume = 247,
	pages = "20004",
}

@misc{goodfellow2014,
      title={Generative Adversarial Networks}, 
      author={Ian J. Goodfellow and Jean Pouget-Abadie and Mehdi Mirza and Bing Xu and David Warde-Farley and Sherjil Ozair and Aaron Courville and Yoshua Bengio},
      year={2014},
      eprint={1406.2661},
      archivePrefix={arXiv},
      primaryClass={stat.ML},
      url={https://arxiv.org/abs/1406.2661}, 
}

@article{ schmidt_2025,
	author = {{Schmidt, Norberto Sebastián} and {Schwab, Alexander} and {Li, Jingjing} and {Rücker, Ulrich} and {Zakalek, Paul} and {Mauerhofer, Eric} and {Dawidowski, Javier} and {Gutberlet, Thomas}},
	title = {Monte Carlo simulations of cold neutron spectra for various para- and ortho-hydrogen ratios using different codes and nuclear data libraries},
	DOI= "10.1140/epjp/s13360-025-06046-0",
	url= "https://doi.org/10.1140/epjp/s13360-025-06046-0",
	journal = {Eur. Phys. J.  Plus},
	year = 2025,
	volume = 140,
	number = 2,
	pages = "114",
}

@article{ HBS2023,
	author = {{Brückel, Thomas} and {Gutberlet, Thomas} and {Baggemann, Johannes} and {Chen, Junyang} and {Claudio-Weber, Tania} and {Ding, Qi} and {El-Barbari, Monia} and {Li, Jingjing} and {Lieutenant, Klaus} and {Mauerhofer, Eric} and {Rücker, Ulrich} and {Schmidt, Norberto} and {Schwab, Alexander} and {Voigt, Jörg} and {Zakalek, Paul} and {Bessler, Yannick} and {Hanslik, Romuald} and {Achten, Richard} and {Löchte, Fynn} and {Strothmann, Mathias} and {Felden, Olaf} and {Gebel, Ralf} and {Lehrach, Andreas} and {Rimmler, Marius} and {Podlech, Holger} and {Meusel, Oliver} and {Ott, Frédéric} and {Menelle, Alain} and {Paulin, Mariano Andrés}},
	title = {The High Brilliance neutron Source (HBS): A project for a next generation neutron research facility},
	DOI= "10.1051/epjconf/202328602003",
	url= "https://doi.org/10.1051/epjconf/202328602003",
	journal = {EPJ Web Conf.},
	year = 2023,
	volume = 286,
	pages = "02003",
}

@Article{2023-Zakalek.,
  author    = {Paul Zakalek and Richard Achten and Johannes Baggemann and Yannick Beßler and Fabian Beule and Thomas Brückel and Junyang Chen and Qi Ding and Monia El-Barbari and Ralf Engels and Olaf Felden and Ralf Gebel and Kirill Grigoryev and Thomas Gutberlet and Romuald Hanslik and Vsevolod Kamerdzhiev and Peter Kämmerling and Harald Kleines and Jingjing Li and Klaus Lieutenant and Fynn Löchte and Eric Mauerhofer and Mariano Andrés Paulin and Ivan Pechenizkiy and Ulrich Rücker and Norberto Schmidt and Alexander Schwab and Alexander Steffens and Fréderic Ott and Yury Valdau and Egor Vezhlev and Jörg Voigt},
  title     = {The High Brilliance Neutron Source Target Stations},
  journal   = {EPJ Web Conf.},
  year      = {2023},
  date      = {2023-10-09},
  volume    = {286},
  pages     = {02004},
  pubstate  = {published},
  url       = {https://doi.org/10.1051/epjconf/202328602004},
  urldate   = {2023-01-01},
  tppubtype = {article},
}

@article{refId15,
title = {The JULIC Neutron Platform, a testbed for HBS},
author = {Paul Zakalek and Johannes Baggemann and Jingjing Li and Ulrich Rücker and Thomas Gutberlet and Thomas Brückel},
url = {https://doi.org/10.1051/epjconf/202429805003},
doi = {10.1051/epjconf/202429805003},
year = {2024},
date = {2024-01-01},
urldate = {2024-01-01},
journal = {EPJ Web Conf.},
volume = {298},
pages = {05003},
pubstate = {published},
tppubtype = {article}
}

@book{Brckel:1016731,
title = {Technical Design Report HBS Volume 2 – Target
Stations and Moderators},
editor = {Thomas Brückel and Thomas Gutberlet},
url = {https://juser.fz-juelich.de/record/1016731},
doi = {10.34734/FZJ-2023-03723},
isbn = {978-3-95806-710-3},
year = {2023},
date = {2023-01-01},
volume = {9-2},
pages = {118},
publisher = {Forschungszentrum Jülich GmbH Zentralbibliothek, Verlag},
address = {Jülich},
series = {Schriften des Forschungszentrums Jülich Reihe Allgemeines
/ General},
pubstate = {published},
tppubtype = {book}
}

@article{coupling2014,
  title={NICE: Non-linear Independent Components Estimation},
  author={Laurent Dinh and David Krueger and Yoshua Bengio},
  journal={arXiv: Learning},
  year={2014},
  url={https://api.semanticscholar.org/CorpusID:13995862}
}

@article{MAF2016,
  title={Improved Variational Inference with Inverse Autoregressive Flow},
  author={Diederik P. Kingma and Tim Salimans and Max Welling},
  journal={ArXiv},
  year={2016},
  volume={abs/1606.04934},
  url={https://api.semanticscholar.org/CorpusID:11514441}
}

@Article{sym16080942,
AUTHOR = {Coccaro, Andrea and Letizia, Marco and Reyes-González, Humberto and Torre, Riccardo},
TITLE = {Comparison of Affine and Rational Quadratic Spline Coupling and Autoregressive Flows through Robust Statistical Tests},
JOURNAL = {Symmetry},
VOLUME = {16},
YEAR = {2024},
NUMBER = {8},
ARTICLE-NUMBER = {942},
URL = {https://www.mdpi.com/2073-8994/16/8/942},
ISSN = {2073-8994},
}

@INPROCEEDINGS{train_vae,
  author={Rivera, Mariano},
  booktitle={2024 IEEE International Conference on Image Processing (ICIP)}, 
  title={How to Train Your VAE}, 
  year={2024},
  volume={},
  number={},
  pages={3882-3888},
  doi={10.1109/ICIP51287.2024.10647557}
}

@article{juwels2019,
  title={JUWELS: Modular Tier-0/1 supercomputer at the J{\"u}lich supercomputing centre},
  author={Krause, Dorian},
  journal={Journal of large-scale research facilities JLSRF},
  volume={5},
  pages={A135--A135},
  year={2019}
}

@inproceedings{sohl2015deep,
  title={Deep unsupervised learning using nonequilibrium thermodynamics},
  author={Sohl-Dickstein, Jascha and Weiss, Eric and Maheswaranathan, Niru and Ganguli, Surya},
  booktitle={International conference on machine learning},
  pages={2256--2265},
  year={2015},
  organization={pmlr}
}

@article{gretton2012kernel,
  title={A kernel two-sample test},
  author={Gretton, Arthur and Borgwardt, Karsten M and Rasch, Malte J and Sch{\"o}lkopf, Bernhard and Smola, Alexander},
  journal={The journal of machine learning research},
  volume={13},
  number={1},
  pages={723--773},
  year={2012},
  publisher={JMLR. org}
}

@inproceedings{karras2019,
  author       = {Tero Karras and
                  Samuli Laine and
                  Timo Aila},
  title        = {A Style-Based Generator Architecture for Generative Adversarial Networks},
  booktitle    = {{IEEE} Conference on Computer Vision and Pattern Recognition, {CVPR}
                  2019, Long Beach, CA, USA, June 16-20, 2019},
  pages        = {4401--4410},
  publisher    = {Computer Vision Foundation / {IEEE}},
  year         = {2019},
  url          = {http://openaccess.thecvf.com/content\_CVPR\_2019/html/Karras\_A\_Style-Based\_Generator\_Architecture\_for\_Generative\_Adversarial\_Networks\_CVPR\_2019\_paper.html},
  doi          = {10.1109/CVPR.2019.00453},
  bibsource    = {dblp computer science bibliography, https://dblp.org}
}

@ARTICLE{genaireview,

  author={Bond-Taylor, Sam and Leach, Adam and Long, Yang and Willcocks, Chris G.},

  journal={IEEE Transactions on Pattern Analysis and Machine Intelligence}, 

  title={Deep Generative Modelling: A Comparative Review of VAEs, GANs, Normalizing Flows, Energy-Based and Autoregressive Models}, 

  year={2022},

  volume={44},

  number={11},

  pages={7327-7347},

  doi={10.1109/TPAMI.2021.3116668}}

\end{document}